# Preprint Traffic Management and Forecasting System Based on 3D GIS

Xiaoming Li[1,2,3], Zhihan Lv[1], Jinxing Hu[1], Baoyun Zhang[4], Ling Yin[1], Chen Zhong[5], Weixi Wang[2,3], Shengzhong Feng[1]
1. Shenzhen Institute of Advanced Technology(SIAT), Chinese Academy of Science, Shenzhen
2. Shenzhen Research Center of Digital City Engineering, Shenzhen, China
3. Key Laboratory of Urban Land Resources Monitoring and Simulation,
Ministry of Land and Resources, Shenzhen, China
4. Jining Institute of Advanced Technology(JIAT), Chinese Academy of Sciences, Jining, China
5. Centre for Advanced Spatial Analysis, University College London, London, England, United Kingdom
Email: lvzhihan@gmail.com

*Abstract*—This paper takes Shenzhen Futian comprehensive transportation junction as the case, and makes use of continuous multiple real-time dynamic traffic information to carry out monitoring and analysis on spatial and temporal distribution of passenger flow under different means of transportation and service capacity of junction from multi-dimensional space-time perspectives such as different period and special period. Virtual reality geographic information system is employed to present the forecasting result.

## I. INTRODUCTION

With increasingly expanded urban size, the modes of road transportation become diversified. The inventory of various motor vehicles keeps sharply increasing, resulting in rapidly increased demands of people for urban traffic and substantial increase in traffic flow. Road infrastructures, traffic control and traditional traffic patterns already can not adapt to development needs of urban traffic. Traffic jam and other issues become more and more serious day by day. At the same time, urbanization also aggravated the issue of urban environmental pollution. At present, the smart traffic application system has been gradually perfected in various fields. However, system information in different fields can not be shared. The collected data failed to give play to its deserved value or provide comprehensive support to the governmental decision making and management, or provide guidance for going out of citizens. Moreover, it is difficult for effective traffic dispersion, resulting in difficulty in giving full play to the effectiveness of traffic infrastructures.

The appearance of new technologies (such as the Internet of Things and cloud computing) brought forth opportunities to the development of smart traffic. Smart traffic is an integrated transportation system effectively integrating various technologies represented by the Internet of Cars and cloud computing (such as intelligent sensing technology, information network technology, communication transmission technology and data processing technology) and applying these technologies into the whole transportation system for the purpose of playing a role in a larger space-time range. Compared with intelligent traffic, smart traffic not only accumulated and passed data but also laid much emphasis on data utilization and exploitation as well as traffic analysis on information, knowledge discovery and decision-making reaction. Moreover, smart technology was used to replace some traditional tasks needing manual identification and determination, so as to achieve the optimization. Additionally, under the development of the Internet of Cars, smart traffic attached much importance to the maximized interconnection and interworking between the traffic information system and other information systems. Thus, it can be seen that further high-efficiency management and deep analysis of traffic data is a key task for the development of smart traffic. It is necessary to summarize and integrate traffic information from different fields and sufficiently exploit the mass information, so as to remit traffic congestion and guarantee traffic safety as well as fast and environmentally-friendly traffic.

The transportation junction refers to the comprehensive facilities established at the junction of two or above lines on one transportation network, or established at the join of several kinds of transportation network, with the functions of transportation organization, transfer, loading and unloading, storage, and information service, etc [10]. The urban comprehensive transportation junction is the important node of urban transportation network, and the efficient management on urban comprehensive transportation junction is the important link to improve urban public traffic system, solve residents transfer, and improve the service quality of public traffic.

As an important part of urban transportation system, the urban passenger terminal is an important joint-point for urban internal traffic and external traffic [44]. The urban passenger terminal often gives a comprehensive consideration on urban external highway passenger transportation and urban public traffic, private transportation, as well as railway, aviation, and other external passenger transportation, in order to establish an organic passenger transportation and bring important benefits to urban development [27]. The reasonable planning and design and efficient management on urban passenger terminal are the important link to improve urban public transportation system, solve residents transfer, and improve the service quality and operation benefits of public transportation [8].

Virtual Reality Geographical Information System (VRGIS) can obtain the landscape geospatial data dynamically, and also perform rich visual 3D analysis and data managements based on Geographical Information System (GIS) data. Accordingly,

'3-D modes' has been proved as a faster decision making tool with fewer errors [29]. A parallel trend, the utilize of bigdata is becoming a hot research topic rapidly recently [1]. GIS data has several characteristics, such as large scale, diverse predictable and real-time, which falls in the range of definition of Big Data [3]. Besides, to improve the accuracy of modeling, the city planning has an increasingly high demand for the realistic display of VR system, however this will inevitably lead to the growth of the volume of data. Virtual scene from a single building to the city scale is also resulting in the increased amount of data. In addition, the concern of usability of WebVRGIS has attracted the attentions to a new challenge, which is fusing all kinds of city information big data by WebVRGIS platform, and thus exploiting the data effectively. Beside massive multi-source spatial data fusing, interaction approach for geo-databases is also expected [2]. Therefore, the management and development of city big data using virtual reality technology is a promising and inspiring approach. Some early related systems from both academy and industry have inspired our work [40] [26] [33] [32]. As a practical tool, most commonly used functions of VRGIS are improved according to practical needs [15].

## II. SYSTEM

The smart traffic cloud service platform is a comprehensive service platform providing timely and abundant traffic information to all categories of traffic participants. The platform was established on the basis of increasingly mature cloud computing technology and the Internet of Things technology. It radically solved disadvantages of the traditional information platform, such as insufficient information processing capability and unsmooth information interaction channel.

The smart traffic cloud service platform is aimed at unified management and mining analysis of mass multi-element isomeric dynamic traffic information, provision of real-time traffic information, improvement to the utilization of traffic information, promotion of the traffic management and travel information service level and provision of supports to traffic management decision making.

Fully integrate traffic resources, systematically process and manage multi-element isomeric traffic data, support interconnection and interworking between traffic data and other information systems, promote mutual collaboration between various traffic functional parts and industrial parts, and promote sharing and the maximum utilization of resources;

Utilize the cloud platform to realize storage and computing of mass traffic information; it not only solved the practical issue of the failure of the current traffic to store mass traffic information but also gradually realized real-time computing and analysis of mass traffic information through strong computing capability of the cloud platform; provide real-time and stable network service to intelligent traffic departments;

In-depth data mining and analysis of multi-source data; it not only enables traffic management personnel and traffic participants to master and understand real-time traffic conditions but also continuously makes comparative analysis on real-time data and historical data, making them to master and understand variation trends and identification abnormities of traffic conditions; upon long-term analysis, it demonstrated for traffic organization and traffic planning in the earlier stage.

The data on smart traffic is sourced from multiple data collection modes: real-time traffic flow information (including speed, flow, occupancy and emergency alarming) of the detected road segment provided by the video detection system, sensibility coil, infrared detection system and others distributed on the road network, floating car data (mainly including records of GPS mounted on taxies, buses and private cars), comprehensively and continuously provided real-time traffic flow information of the whole traffic network, car state information, pavement condition information, road acoustic environment and road air quality provided by the information collection module at the front end of the Internet of Cars, bidirectional transmission information of multiple levels (for instance, car car, car roadside equipment and car information platform) collected by the platform of the Internet of Cars. All these information collection means jointly generated mass multi-element isomeric dynamic traffic information.

The transportation system is a continuously dynamic complex giant system. The development of information technology makes it possible for us to continuously observe this complex system and record its dynamic process. The floating car data (mainly including GPS positioning record installed on taxis, buses, and private cars) which has rapid development in recent years provides the real-time traffic flow information of whole transportation network in a more comprehensive and continuous way [30]; the personal mobile phone positioning data under exploration has great potential in getting a detailed knowledge of residents trip characteristics and real-time detection of traffic flow [36]. Those information acquisition means produce massive real-time dynamic traffic information.

The real-time dynamic traffic information includes two features, that is, real-time and dynamic. The real-time refers to immediate acquisition, processing, and release of information, and the dynamic refers to the situation that the acquisition, processing, and release of traffic information continuously change with traffic condition on one hand, and the comparison and analysis of historical data are continuously made on the other hand. There is no doubt that the intelligent transportation system is the only road for modern society to improve traffic environment. Besides, the acquisition, analysis, and release of real-time dynamic traffic information are a key technology and basis to develop intelligent transportation system.

In this work, we utilize Futian comprehensive transportation as a convincing case to present WebVRGIS [22], which is based on WebVR render engine [25]. Shenzhen is a thirty-years new city, however, it has the highest population density in China, which reaches 7785 people per square kilometer (2013). It causes some embarrassments to the city information management [14]. While virtual environments have proven to significantly improve public understanding of 3D planning data [6]. To share of information resources of all departments and the dynamic tracking for the geospatial information of population and companies, by construction an integrated information platform of social services. The use of virtual reality as visual means changed the traditional image of the city [7].

The geographic statistical analysis is to assist management decision-making and conduct data analysis.

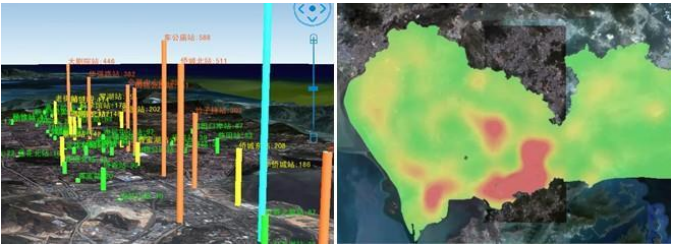

Fig. 1. Left: Passenger flow forecast of various stations; Right: Population distribution

The 2D statistical analysis visualization is overlapped with a white background on 3D virtual reality environment, since it's more intuitive and a cognitively less demanding display system, which lessens the cognitive workload of the user [29].

The innovation points of this work include, real-time dynamic comprehensive transportation data mining, three-dimensional GIS analysis and release as for transportation junction; Carry out comprehensive assessment on service scope of junction through analyzing long-term dynamic traffic data; Carry out comprehensive analysis on actual travel time and then comprehensive assessment on service capacity of public transportation (accessibility and reliability).

Spatiotemporal database model and visualization has been considered in the design of the system [43].

### A. Traffic basic information management subsystem

This subsystem is used for centralized management of traffic-related urban space data resources and establishment of multi-scale and multi-resolution space information infrastructure databases, including the vector database, satellite image database, 3D model database and traffic facility database of urban road network. Under the support from the 3D traffic geographic information system, this subsystem provides various functions, such as map visualization, information inquiry and information updating of various types of traffic basic information.

### B. Dynamic traffic information processing subsystem

This subsystem is connected with the traffic information collection port. On the one hand, this subsystem is used to provide data acquisition and preprocessing functions of dynamic traffic data (such as loading, updating, converting and fault-tolerance processing). On the other hand, it is used to provide various models for data integration, for instance, weight fusion, historical analysis, time-space complementation, microcosmic traffic flow transmission model and regional network traffic flow analysis model.

### C. Dynamic traffic network analysis subsystem

This subsystem is used for traffic flow analysis and forecasting of the whole network on the basis of output results from the dynamic traffic information processing system. Its functions include analysis on congestion conditions of various road segments, judgment on the severity of traffic events, forecasting of time-space impact scope of traffic events, forecasting of modes of transportation of the whole network after several minutes or hours (including road traffic, rail traffic and common public transportation) and designating of real-time accessibility of network nodes.

### D. Planning decision making auxiliary subsystem

This subsystem is used to provide statistic analysis on long-term traffic flow and traffic events of appointed network nodes (such as daily car/ passenger flow, different-period car/ passenger flow and traffic accident statistics of a certain road segment or traffic hub), so as to make it convenient for supervision departments to understand long-term conditions of the whole traffic system or some important nodes and take this as the basis for decision making. Besides, time-space trace data of each car in the road network provided by the Internet of Cars is utilized to form a continuous and comprehensive motor vehicle travel OD database. This database is used in combination with card data on public transportation, so as to provide a comprehensive travel OD database about modes of transportation of citizens. On this basis, this system provides the OD analysis function of various modes of transportation, providing important decision making supports to traffic planning. Additionally, this system is used for analysis on travel behaviors of population, taxi path, incidence degree of public transportation and road travel speed under special weather conditions (such as rainstorm, typhoon and heavy fog) and in special periods (such as festivals and holidays and important events), so as to provide the analysis and forecasting function of abnormal traffic flow.

### E. 3D traffic geographic information subsystem

This subsystem is used for 3D visualization of spatial information and data analysis results. It is of basic functions of the 3D geographic information system, including selection of object, various operations, setting of time and sun, underground mode, coverage control, horizontal distance measurement, spatial distance measurement, vertical distance measurement, area measurement, setting of navigation mode, setting of 3D mode, setting of indoor navigation mode, setting of eagle-eye map and landform video. In addition, this system also supports hierarchical 3D visualization expression & query of multi-scale traffic network flow congestion conditions, 3D visualization expression & query of various traffic events, visualization of decision-making plan and traffic analysis model based on 3D space data model.

## III. METRO PASSENGER FLOW FORECASTING

The passenger flow forecasting refers to the index which reflects the demand characteristics of traffic passenger flow via forecasting the cross sectional flow of urban transport lines within certain period and inter-station OD. Upon planning the traffic network, the passenger flow analysis result of different traffic network schemes is the main content based on which the line network is selected.

Peoples travel demand is the unity of randomness and regularity, and it is reflected in large quantity of statistical data. Therefore, it is difficult to accurately forecast the passenger flow in one day; however, the forecasting can be made within certain confidence, which is also the starting point of forecasting.

TABLE I. RESULTS OF REPEATABLE DOUBLE FACTOR VARIANCE ANALYSIS OF OUTBOUND TRAFFICS FROM MONDAY TO SUNDAY

| Source of Difference | SS | df | MS | F | P-value | F crit |
|---|---|---|---|---|---|---|
| Sample | 8766084 | 6 | 1461014 | 252.5877 | 2.49E-81 | 2.152911 |
| Row | 71310564 | 7 | 10187223 | 1761.22 | 1.2E-153 | 2.06446 |
| Interactive | 20207067 | 42 | 481120.6 | 83.17865 | 1.91E-92 | 1.457365 |
| Internal | 971743 | 168 | 5784.185 | | | |
| Total | 1.01E+08 | 223 | | | | |

TABLE II. REGRESSION MODELING RESULTS OF OUTBOUND TRAFFICS AT TIME PERIOD FROM MONDAY TO THURSDAY REGRESSION STATISTICAL RESULTS

| Regression | |
|---|---|
| Multiple R | 0.994189108 |
| R Square | 0.988411983 |
| Adjusted R Square | 0.987730335 |
| Standard Error | 81.77353609 |
| Observed Value | 127 |

TABLE III. REGRESSION MODELING RESULTS OF OUTBOUND TRAFFICS AT TIME PERIOD FROM MONDAY TO THURSDAY VARIANCE ANALYSIS RESULTS

| | df | SS | MS | F | Significance F |
|---|---|---|---|---|---|
| Regression Analysis | 7 | 67873681 | 9696240 | 1450.033 | 5.5236E-112 |
| Residual Error | 119 | 795742.4 | 6686.911 | | |
| Total | 126 | 68669424 | | | |

There are many kinds of traffic passenger flow forecasting models, and the common models include regression forecasting model and time series prediction model [13] [16] [12] [28]. In the observation period of this research, the metro passenger flow shows strong regularity and stability without long-term change trend; therefore, this research adopts regression forecasting model to forecast the metro passenger flow. While we forecast the passenger flow, there may be many independent variables which have influence on the result of dependent variable (passenger flow), but the actual situation is that it is only allowed to find out several independent variables which have important influence on the dependent variable and ignore other independent variables. In specific application, it is required to screen out some main independent variables which cause influence on the result of dependent variable for research and analysis, and the analysis of variance theory is applied in this screen-out process.

The analysis of variance is a kind of statistical analysis method in which the analysis and processing is made for significance of difference in mean values of some sets of experimental data. The passenger flow samples are different in different periods in each day; therefore, it is able to know that the date and period are 2 variables which influence passenger flow; through dual-factor analysis of variance, it is able to determine whether these two factors are important factors which influence passenger flow and then regard them as input variables in later forecasting analysis if so.

Due to difference in inbound and outbound metro passenger flow, this research carries out forecasting modeling and model validation on inbound and outbound passenger flow respectively.

For outbound traffic, predictive modeling and model validation contain the following steps:

1. Outbound traffics of four weeks with normal data at each time period in July and August are selected as modeling samples.

2. All modeling samples are of repeatable double factor variance analysis in terms of day of week and time period, and results are as follows:

According to Table I, the sample group has $p-value = 2.49E-81 \gg 0.05$, and thus its traffics have significant difference, i.e. outbound traffics at different time periods from Monday to Sunday have significant difference; traffic of each time period within the group has difference, traffic $\gg 0.05$, and so there is also significant difference among outbound traffics at all time periods within the group. Thus, both day of week and time period can be considered as influencing factors of outbound traffic.

3. To further test differences in traffic changes within a week, this study carries out repeatedly combined double factor variance analysis and finds: no significant difference in outbound traffics at time period from Monday to Thursday; and significant difference in outbound traffics at time periods from Monday to Thursday, on Friday, Saturday and Sunday.

4. Based on the above analysis, regression models are established with respect to outbound traffics at different time periods from Monday to Thursday, on Friday, Saturday and Sunday. This study divides time period into 8 parts, and makes time period variable become a categorical independent variable with 8 types, and thus 7 dependent variables are generated. These four regression models are expressed as follows:

(1) Outbound traffics at time period from Monday to Thursday: $Flow\ out\ Mon = a_1t_1 + a_2t_2 + a_3t_3 + a_4t_4 + a_5t_5 + a_6t_6 + a_7t_7 + a_8$

(2) Outbound traffics on Friday: $Flow\ out\ Fri = b_1t_1 + b_2t_2 + b_3t_3 + b_4t_4 + b_5t_5 + b_6t_6 + b_7t_7 + b_8$

(3) Outbound traffics on Saturday: $Flow\ out\ Sat = c_1t_1 + c_2t_2 + c_3t_3 + c_4t_4 + c_5t_5 + c_6t_6 + c_7t_7 + c_8$

(4) Outbound traffics on Sunday: $Flow\ out\ Sun = d_1t_1 + d_2t_2 + d_3t_3 + d_4t_4 + d_5t_5 + d_6t_6 + d_7t_7 + d_8$

Where, $t_1, t_2, \ldots t_7$ are dependent variables of 8 time periods. When $t_1 = 1, t_2 = t_3 = \ldots = t_7 = 0$, dependent variables denote outbound traffics of the first time period; when $t_2 = 1, t_1 = t_3 = \ldots = t_7 = 0$, dependent variables denote outbound traffics of the second time period; reason by analogy; when $t_1 = t_2 = t_3 = \ldots = t_7 = 0$, dependent variables denote outbound traffics of the eighth time period. $a_i, b_i, c_i, d_i$ are independent variable parameters and constant terms of these four models respectively.

5. By modeling with sample data in Table II, results of predictive model of outbound traffics at time period from Monday to Thursday are as follows:

According to Table II, predictive model of outbound traffics at time period from Monday to Thursday has Adjusted R Square=98.77%, i.e. this model can explain 98.77% of sample data; the entire model has statistical significance at level $\alpha = 0.05$; all parameters of the model have p-value smaller than 0.05, which means that all of these parameters have statistical significance at level $\alpha = 0.05$.

TABLE IV. REGRESSION MODELING RESULTS OF OUTBOUND TRAFFICS AT TIME PERIOD FROM MONDAY TO THURSDAY MODEL COEFFICIENT RESULTS

|   | Coefficients | Standard Error | t Stat | P-value |
|---|---|---|---|---|
| Intercept | 522.6875 | 20.44338 | 25.56756 | 3.61E-50 |
| 1 | 1963.579167 | 29.38922 | 66.81291 | 3.36E-96 |
| 0 | 1014.8125 | 28.91131 | 35.10088 | 1.27E-64 |
| 0 | 676.375 | 28.91131 | 23.39482 | 2.46E-46 |
| 0 | 776.125 | 28.91131 | 26.84503 | 2.55E-52 |
| 0 | 1124.0625 | 28.91131 | 38.87968 | 1.74E-69 |
| 0 | 2476.875 | 28.91131 | 85.67149 | 8.7E-109 |
| 0 | 717.4375 | 28.91131 | 24.81511 | 7.23E-49 |

Similarly, modeling results of outbound traffics at time periods of Friday, Saturday and Sunday are as follows: predictive model of outbound traffics on Friday has Adjusted R Square=98.62% and statistical significance at level $\alpha = 0.05$, and all of its parameters have p-value smaller than 0.05; predictive model of outbound traffics on Saturday has Adjusted R Square=91.43% and statistical significance at level $\alpha = 0.05$, and all of its parameters have p-value smaller than 0.05; predictive model of outbound traffics on Sunday has Adjusted R Square=94.41% and statistical significance at level $\alpha = 0.05$, and all of its parameters have p-value smaller than 0.05.

To sum up, predictive models of outbound traffics at four time periods from Monday to Sunday have good imitative effects of sample data and are expressed as follows:

(1) Outbound traffics at time period from Monday to Thursday: $Flow\ out\ Mon = 1963.58t_1 + 1014.81t_2 + 676.38t_3 + 776.13t_4 + 1124.06t_5 + 2476.88t_6 + 717.44t_7 + 522.69$

(2) Outbound traffics on Friday: $Flow\ out\ Fri = 1711.83t_1 + 826t_2 + 591t_3 + 697.75t_4 + 949t_5 + 2660.25t_6 + 905.5t_7 + 641.5$

(3) Outbound traffics on Saturday: $Flow\ out\ Sat = 609.67t_1 + 692.5t_2 + 714.5t_3 + 825.75t_4 + 707.75t_5 + 869.25t_6 + 482.75t_7 + 616$

(4) Outbound traffics on Sunday: $Flow\ out\ Sun = 208.92t_1 + 528.75t_2 + 578.75t_3 + 694.5t_4 + 747t_5 + 821.75t_6 + 585.5t_7 + 653.75$

Outbound traffics at time periods with normal data within a week of August are selected as model validation samples, and such data includes:

Validate prediction formula with validation samples and calculate the accuracy of each independent predicted value with absolute percentage error (APE): $APE = 100\% \times \|predictedvalue - actualvalue\|/actualvalue$. Validation results are as follows:

As can be known from Table IV, the mean error of predictive model of subway outbound traffics at different time periods is 7.88%, and thus this model has good validation effects.

## IV. GRAPHIC USER INTERFACE

With a 3D earth model as the browser, this system is loaded with all 3D model data and the 3D visualization analysis result. By zooming in the camera horizon with 'Zoom in' function, it is possible to see the detailed level of buildings. By selecting the house inquiry, it is possible to find out the information of the address and owner, and to locate the house in the 3D scene. 3D roaming function can not only conduct soaring top view observation above the virtual community [18] [41], but can also observe the detailed layout near the street, and further enter the building to observe the internal building structure. In addition to the observation of above-ground model, it can also observe and manage the underground data under the city ground.

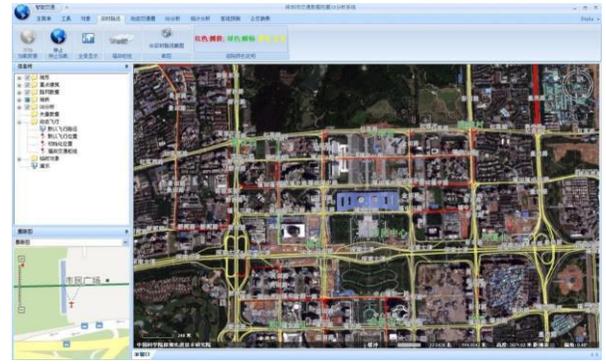

Fig. 2. Local Map of Real-time Traffic Analysis and Distribution

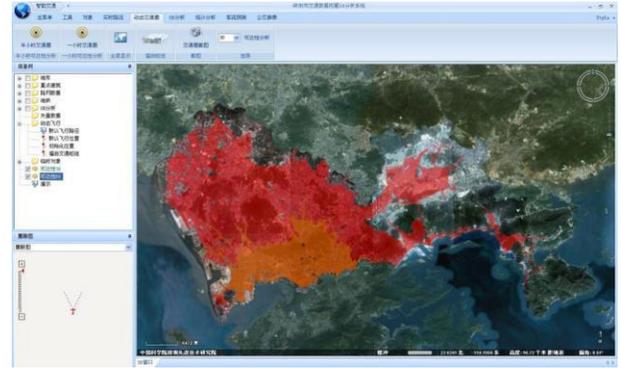

Fig. 3. Loop Map of Real-time Dynamic Traffic 30 and 60 Minutes

With Futian comprehensive transportation junction with the function of a car harbor in the earliest time in Shenzhen, this research utilized continuous multi-element real-time dynamic traffic information (including data on taxi and floating car, card data on public transportation and information about long-distance passenger transportation) to monitor and analyze the passenger flow spatial and temporal distribution and the service capability of different modes of transportation in the junction in different time frames and special periods. Moreover, multi-element real-time dynamic traffic information was subject to management and analysis of the 3D geographic information system, providing technical supports and case reference to the construction of the urban traffic data management and analysis system.

Several key issues have been solved as followed.

## V. TRAFFIC INFORMATION FAULT-TOLERANCE PROCESSING

Due to work breakdown, deviation and other reasons of traffic detector and transmission devices, traffic data inevitably involves error, missing and other problems. Therefore, the platform is used to detect the original data, remove abnormal data and repair incomplete data, guaranteeing data integrity and correctness.

Fault-tolerance processing mainly includes judgment on obliterated data, identification for abnormal data and repair of incomplete data. Obliterated data is mainly judged by continuously scanning dynamic data in a certain period of time according to set data collection time and format. In the abnormal data identification process, the abnormal data is distinguished whether it belongs to equipment failure data or correct traffic abnormal data caused by traffic events, abnormal weather or other reasons. For repair of incomplete data, mathematical model is used for analysis and forecasting, so as to supplement the incomplete data.

## VI. Multi-element isomeric traffic data integration

Collection modes of traffic data include floating car, camera, radar, microwave and manual labor. Data collected from different sources is different in structure, accuracy, position and time. As a result, multi-element isomeric traffic data must be subject to data integration before traffic analysis. With the development of the Internet of Cars and other intelligent sensing technologies and network communication technology, the multi-element isomeric feature of traffic data is highlighted day by day, resulting in greater challenges to data integration.

Various types of data are taken full advantage of by improving data accuracy and the network coverage of data. On the one hand, space data is subject to complementation, mutual verification and superposition computing through various data models and statistical models, so as to process multi-source data and form more comprehensive traffic description. On the other hand, with dispersed cars as basic description unit, the collected road segment is subject to more accurate data integration from a microcosmic perspective, so as to obtain more accurate traffic information.

## VII. Mass traffic information storage and computing

Traffic data is of huge capacity, diversified sources and frequent updating. It is the important technical guarantee for the construction of smart traffic to effectively store and manage this mass data and make it meet the requirements of the traffic system application for high availability and high reliability. For storage of mass data, it is necessary to establish a database featured by rapid and convenient query and management and realize high-efficiency data transmission under limited network bandwidth. At the same time, it needs supports from a high-performance computer platform for acquisition, integration, analysis and application of mass traffic data, so as to analyze and forecast dynamic traffic flow in real time.

Real-time analysis and forecasting of traffic flow is always a difficulty in traffic analysis. Changes in traffic network are rapid and complicated. With the improvement to the updating speed of dynamic traffic data, there are higher and higher requirements for real-time analysis and forecasting of the traffic capacity of various modes of transportation. On the basis of data integration, this platform is used to analyze the congestion degree of the road network, judge the severity of traffic events, forecast the time-space impact scope of traffic events and forecast the road segment speed of various modes of transportation in the whole network after several minutes or hours.

## VIII. Conclusion

With the development of 3D geographic information system and peoples direct demands for 3D scene, it is an inexorable trend to establish a 3D traffic geographic information system. This platform supports hierarchical 3D visualization expression & query of multi-scale traffic network flow, 3D visualization expression & query of various traffic events, visualization of decision-making plan and research on spatial analysis model based on 3D space data model.

This platform utilizes virtualization, distributed computation and cloud technology to guarantee safe and stable operation of the system, improving security and disaster tolerance of data and the system to the maximum extent. Based on the system framework of cloud computing, this platform broke through the performance bottleneck of the traditional system, realizing real-time detecting and intelligent monitoring of traffic information and easy processing of mass data, and meeting the high performance needs of mass dynamic traffic information for real-time processing. It is characterized with strong application load adaptability, wide compatibility and easy expansion. This platform is compatible with multi-channel information distribution modes (including PSTN, Internet and 3G wireless) and supports such information distribution terminals as PC, PDA, mobile phone and tablet PC.

## IX. Future Work

Through long-term monitoring and analysis, the long-term transportation junction demand model, and long-term passenger flow forecasting and early-warning model are established under the condition of combing with economic development and urban planning. The analysis is made on population travel behavior, taxi route, degree of influence on public bus, and road travelling speed under special weather conditions (such as rainstorm, typhoon, and heavy fog, etc.) The deeper data mining is made, such as emergency evacuation aided decision support, monitoring and forecasting on large-scale group event, assisting crowd and vehicle evacuation under emergency. The system also support ocean data visualiza- tion [31] [23] [24], which has potential to extend to ocean traffic forcasting system. The separated core technology of our system also has potential to be applied into other fields, such as climate [34], biology [35], clinical assist [19]. Some novel interaction approaches are considered to be integrated in our future work [17] [21] [20]. The new network data management algorithm [37] [38] [11], smart grid system [5] [4], data classification method [9], pedestrian detector technology [39] and stereoscopic 3D visualization approach [42] will be also considered.

## Acknowledgments

The authors are thankful to the National Natural Science Fund for the Youth of China (41301439) and Electricity 863 project(SS2015AA050201).

## REFERENCES

[1] Big data specials. *Nature*, 455(7209), Sept. 2008.

[2] M. Breunig and S. Zlatanova. Review: 3d geo-database research: Retrospective and future directions. *Comput. Geosci.*, 37(7):791–803, July 2011.

[3] F. Briggs. Large data - great opportunities. Presented at IDF2012, Beijing, 2012.

[4] L. Che, M. Khodayar, and M. Shahidehpour. Adaptive protection system for microgrids: Protection practices of a functional microgrid system. *IEEE Electrification Magazine*, 2(1):66–80, 2014.

[5] L. Che and M. Shahidehpour. Dc microgrids: Economic operation and enhancement of resilience by hierarchical control. *IEEE Transactions Smart Grid*, 5(5):2517–2526, 2014.

[6] E. Chow, A. Hammad, and P. Gauthier. Multi-touch screens for navigating 3d virtual environments in participatory urban planning. In *CHI '11 Extended Abstracts on Human Factors in Computing Systems*, CHI EA '11, pages 2395–2400, New York, NY, USA, 2011. ACM.

[7] C. DiSalvo and J. Vertesi. Imaging the city: Exploring the practices and technologies of representing the urban environment in hci. In *CHI '07 Extended Abstracts on Human Factors in Computing Systems*, CHI EA '07, pages 2829–2832, New York, NY, USA, 2007. ACM.

[8] A. Etches, D. Parker, S. Ince, and P. James. Utis (urban transportation information system) a geo-spatial transport database. In *Proceedings of the 8th ACM International Symposium on Advances in Geographic Information Systems*, GIS '00, pages 83–88, New York, NY, USA, 2000. ACM.

[9] H. Fei and J. Huan. Structured sparse boosting for graph classification. *ACM Transactions on Knowledge Discovery from Data (TKDD)*, 9(1):4, 2014.

[10] Z. Huanyu, X. Liangjie, and L. Jun. Research on service level of city railway transport hub. In *Proceedings of the 5th International Conference on Wireless Communications, Networking and Mobile Computing*, WiCOM'09, pages 4290–4293, Piscataway, NJ, USA, 2009. IEEE Press.

[11] D. Jiang, Z. Xu, P. Zhang, and T. Zhu. A transform domain-based anomaly detection approach to network-wide traffic. *Journal of Network and Computer Applications*, 40:292–306, 2014.

[12] M. Khashei, S. Reza Hejazi, and M. Bijari. A new hybrid artificial neural networks and fuzzy regression model for time series forecasting. *Fuzzy Sets Syst.*, 159(7):769–786, Apr. 2008.

[13] H.-Z. Li, S. Guo, C.-J. Li, and J.-Q. Sun. A hybrid annual power load forecasting model based on generalized regression neural network with fruit fly optimization algorithm. *Know.-Based Syst.*, 37:378–387, Jan. 2013.

[14] X. Li, Z. Lv, B. Zhang, W. Wang, S. Feng, and J. Hu. Webvrgis based city bigdata 3d visualization and analysis. In *Visualization Symposium (PacificVis), 2015 IEEE Pacific*. IEEE, 2015.

[15] H. Lin, M. Chen, G. Lu, Q. Zhu, J. Gong, X. You, Y. Wen, B. Xu, and M. Hu. Virtual geographic environments (vges): A new generation of geographic analysis tool. *Earth-Science Reviews*, 126(0):74 – 84, 2013.

[16] K.-P. Lin, P.-F. Pai, Y.-M. Lu, and P.-T. Chang. Revenue forecasting using a least-squares support vector regression model in a fuzzy environment. *Inf. Sci.*, 220:196–209, Jan. 2013.

[17] Z. Lv. Wearable smartphone: Wearable hybrid framework for hand and foot gesture interaction on smartphone. In *2013 IEEE International Conference on Computer Vision Workshops*, pages 436–443. IEEE, 2013.

[18] Z. Lv, G. Chen, C. Zhong, Y. Han, and Y. Y. Qi. A framework for multi-dimensional webgis based interactive online virtual community. *Advanced Science Letters*, 7(1):215–219, 2012.

[19] Z. Lv, C. Esteve, J. Chirivella, and P. Gagliardo. A game based assistive tool for rehabilitation of dysphonic patients. In *Virtual and Augmented Assistive Technology (VAAT), 2015 3nd Workshop on*. IEEE, 2015.

[20] Z. Lv, L. Feng, S. Feng, and H. Li. Extending touch-less interaction on vision based wearable device. In *Virtual Reality (VR), 2015 iEEE*. IEEE, 2015.

[21] Z. Lv, L. Feng, H. Li, and S. Feng. Hand-free motion interaction on google glass. In *SIGGRAPH Asia 2014 Mobile Graphics and Interactive Applications*. ACM, 2014.

[22] Z. Lv, S. Rhman, and G. Chen. Webvrgis: A p2p network engine for vr data and gis analysis. In M. Lee, A. Hirose, Z.-G. Hou, and R. Kil, editors, *Neural Information Processing*, volume 8226 of *Lecture Notes in Computer Science*, pages 503–510. Springer Berlin Heidelberg, 2013.

[23] Z. Lv and T. Su. 3d seabed modeling and visualization on ubiquitous context. In *SIGGRAPH Asia 2014 Posters*, page 33. ACM, 2014.

[24] Z. Lv, T. Su, X. Li, and S. Feng. 3d visual analysis of seabed on smartphone. In *Visualization Symposium (PacificVis), 2015 IEEE Pacific*. IEEE, 2015.

[25] Z. Lv, T. Yin, Y. Han, Y. Chen, and G. Chen. Webvr–web virtual reality engine based on p2p network. *Journal of Networks*, 6(7), 2011.

[26] C. Ma, G. Chen, Y. Han, Y. Qi, and Y. Chen. An integrated vr–gis navigation platform for city-region simulation. *Comput. Animat. Virtual Worlds*, 21(5):499–507, Sept. 2010.

[27] H. J. Miller and Y.-H. Wu. Gis software for measuring space-time accessibility in transportation planning and analysis. *Geoinformatica*, 4(2):141–159, June 2000.

[28] S.-K. Oh, M.-S. Kim, T.-D. Eom, and J.-J. Lee. Heterogeneous local model networks for time series prediction. *Appl. Math. Comput.*, 168(1):164–177, Sept. 2005.

[29] T. Porathe and J. Prison. Design of human-map system interaction. In *CHI '08 Extended Abstracts on Human Factors in Computing Systems*, CHI EA '08, pages 2859–2864, New York, NY, USA, 2008. ACM.

[30] M. A. Quddus, W. Y. Ochieng, and R. B. Noland. Current map-matching algorithms for transport applications: State-of-the art and future research directions. *Transportation Research Part C: Emerging Technologies*, 15(5):312 – 328, 2007.

[31] T. Su, Z. Lv, S. Gao, X. Li, and H. Lv. 3d seabed: 3d modeling and visualization platform for the seabed. In *Multimedia and Expo Workshops (ICMEW), 2014 IEEE International Conference on*, pages 1–6. IEEE, 2014.

[32] J. Tan, F. Deng, et al. Design and key technology of urban landscape 3d visualization system. *Procedia Environmental Sciences*, 10:1238–1243, 2011.

[33] J. Tan, X. Fan, and Y. Ren. Methodology for geographical data evolution: three-dimensional particle-based real-time snow simulation with remote-sensing data. *Journal of Applied Remote Sensing*, 8(1):084598–084598, 2014.

[34] Y. Tang, S. Zhong, L. Luo, X. Bian, W. E. Heilman, and J. Winkler. The potential impact of regional climate change on fire weather in the united states. *Annals of the Association of American Geographers*, 105(1):1–21, 2015.

[35] A. Tek, B. Laurent, M. Piuzzi, Z. Lu, M. Chavent, M. Baaden, O. Delalande, C. Martin, L. Piccinali, B. Katz, et al. Advances in human-protein interaction-interactive and immersive molecular simulations. *Biochemistry, Genetics and Molecular Biology"Protein-Protein Interactions-Computational and Experimental Tools'*, pages 27–65, 2012.

[36] A. Vaccari, F. Calabrese, B. Liu, and C. Ratti. Towards the socioscope: An information system for the study of social dynamics through digital traces. In *Proceedings of the 17th ACM SIGSPATIAL International Conference on Advances in Geographic Information Systems*, GIS '09, pages 52–61, New York, NY, USA, 2009. ACM.

[37] Y. Wang, W. Jiang, and G. Agrawal. Scimate: A novel mapreduce-like framework for multiple scientific data formats. In *Cluster, Cloud and Grid Computing (CCGrid), 2012 12th IEEE/ACM International Symposium on*, pages 443–450. IEEE, 2012.

[38] Y. Wang, A. Nandi, and G. Agrawal. Saga: Array storage as a db with support for structural aggregations. In *Proceedings of the 26th International Conference on Scientific and Statistical Database Management*, SSDBM '14, pages 9:1–9:12, New York, NY, USA, 2014. ACM.

[39] J. Xu, D. Vazquez, A. M. Lopez, J. Marin, and D. Ponsa. Learning a part-based pedestrian detector in a virtual world. 2014.

[40] J. Zhang, J. Gong, H. Lin, G. Wang, J. Huang, J. Zhu, B. Xu, and J. Teng. Design and development of distributed virtual geographic environment system based on web services. *Inf. Sci.*, 177(19):3968–3980, Oct. 2007.

[41] M. Zhang, Z. Lv, X. Zhang, G. Chen, and K. Zhang. Research and application of the 3d virtual community based on webvr and ria. *Computer and Information Science*, 2(1):P84, 2009.

[42] Y. Zhang, G. Jiang, M. Yu, and K. Chen. Stereoscopic visual attention model for 3d video. In *Advances in Multimedia Modeling*, pages 314–324. Springer Berlin Heidelberg, 2010.

[43] C. Zhong, T. Wang, W. Zeng, and S. M. Arisona. Spatiotemporal visualisation: A survey and outlook. In *Digital Urban Modeling and Simulation*, pages 299–317. Springer, 2012.

[44] X. Zhou, D. Di, X. Yang, and D. Wu. Location optimization of urban passenger transportation terminal. In *Proceedings of the 2010 International Conference on Optoelectronics and Image Processing - Volume 01*, ICOIP '10, pages 668–671, Washington, DC, USA, 2010. IEEE Computer Society.